\documentclass[11pt]{article}
\usepackage{epsfig}

\begin{document}

\title{Inferring the success parameter $p$ 
of a binomial model from small samples affected  by background}
\author{G. D'Agostini}

\date{}

\maketitle

\begin{abstract}
The problem of inferring the binomial parameter $p$
from  $x$ {\it successes}  obtained in $n$ {\it trials}
is reviewed and extended to take into account the presence 
of background, that can affect
the data in two ways: {\it a}) fake successes are due to 
a background modeled as a Poisson process of known intensity; 
{\it b}) fake trials are  due to 
a background modeled as a Poisson process of known intensity,
each trial being characterized by a known 
success probability $p_b$.
\end{abstract}

\section{Introduction}
An important class of experiments consists in counting
`objects'. In fact, we are often interested in measuring their
{\it density}  in time, space, or  both (here `density' 
stands for a general term, that in the domain of time is
equivalent to `rate')  
or the {\it  proportion} of those objects 
that have a certain character in common.
For example, particle physicists
might be interested in cross sections and
branching ratios, astronomers in density of galaxies 
in a region of the sky
or in the ratio of galaxies exhibiting some special features.

A well known problem in counting experiments 
is that we are rarely in the ideal situation of being able to 
count individually and at a given time all the objects
of interest. More often we have to rely
an a sample of them. Other 
problems that occur  in real environments,
especially in frontier research, are detector inefficiency
and presence of background: sometimes we lose objects
in counting; other times we might be confused by other
objects that do not belong to the classes we are looking for,
though they are observationally indistinguishable from
the objects of interest. 

We focus here on the effect of background in measurements of
proportions. For a extensive treatment of the effect of background
on rates, i.e. measuring the intensity of a Poisson process
in presence of background, see Ref.~\cite{RPoisson}, as well as
chapters 7 and 13 of Ref.~\cite{BR}. 

The paper is structured as follows. 
In section \ref{sec:binomial} we introduce the 
`direct' and `inverse'
probabilistic problems related to the binomial distribution
and the two cases of background that will be considered.
In section \ref{sec:no_bkgd} we go through the standard
text-book case in which background is absent, but we
discuss also, in some depth, the issue of 
how prior knowledge does or does not influence the probabilistic
conclusions.
Then, in the following two sections we come to the
specific issue of this paper, and finally the paper ends 
with the customary short conclusions. 

\section{The binomial distribution and its inverse problem}
\label{sec:binomial}
An important class of counting experiments can be modeled
as independent Bernoulli trials. In each trial 
we believe that a {\it success} will occur with probability 
$p$, and a  {\it failure}
with probability $q=1-p$. 
If we consider $n$ independent trials, all with the same 
probability $p$, we might be interested in the total
number of successes, independently of their order. 
The total number of successes $X$ can range between $0$ and $n$,
and our belief on the outcome $X=x$ can be evaluated
 from the probability of each success
and some combinatorics. The result is the well known
{\it binomial} distribution, hereafter indicated with ${\cal B}_{n,p}$:
\begin{equation}
f(x\,|\,{\cal B}_{n,p}) =
\frac{n!}{(n-x)!\,x!}\, p^x\, (1-p)^{n-x} \, ,
\hspace{1.0 cm}
\left\{ \begin{array}{l}   n = 1, 2, \ldots,  \infty \\
                           0 \le p \le 1 \\
                           x = 0, 1, \ldots, n \end{array}\right.\,,
\label{eq:binomial}
\end{equation} 
having {\it expected value} and {\it standard deviation}
\begin{eqnarray}
\mbox{E}(X) &  = &  n\,p \\
\sigma(x)   & = & \sqrt{n\,p\, (1-p)}\,.
\end{eqnarray} 
We associate the formal quantities  expected value 
and   standard deviation to the concepts of 
(probabilistic) {\it prevision}
and {\it standard uncertainty}. 

The binomial distribution describes what is sometimes called 
a {\it direct probability} problem, i.e. calculate 
the probability of the experimental outcome $x$ (the {\it effect})
given $n$ and an assumed value of $p$. The {\it inverse} 
problem is what concerns mostly scientists: {\it infer $p$ given
$n$ and $x$}. In probabilistic terms, we are 
interested in $f(p\,|\,n,x)$.
 Probability inversions 
are performed, within probability theory, using Bayes theorem,
that in this case reads
\begin{eqnarray}
f(p\,|\,x,n,{\cal B}) & \propto & f(x\,|\,{\cal B}_{n,p}) \cdot f_\circ(p)\,
\label{eq:inf_binom}
\end{eqnarray}
where $f_\circ(p)$ is the {\it prior}, $f(p\,|\,x,n,{\cal B})$
the {\it posterior}
(or {\it final}) and $ f(x\,|\,{\cal B}_{n,p})$ 
the {\it likelihood}. 
The proportionality factor is calculated
from normalization. [Note the use of $f(\cdot)$ for the 
several probability functions as well as probability density functions
(pdf), also within the same formula.]
The solution of
Eq.~(\ref{eq:inf_binom}), related to the names of Bayes and Laplace,
is presently a kind of first text book exercise in the so called 
Bayesian inference (see e.g.  Ref.~\cite{BR,RPP}).
The issue of priors in this kind of problems will be discussed
in detail in  Sec.~\ref{ss:priors}, especially for the critical
cases of $x=0$ and $x=n$. 

The problem can be complicated by the presence of background.
This is the main subject of this paper, and we shall focus on
two kinds of background.
\begin{enumerate}
\item[{\it a)}]
{\bf Background can  only affect ${\mathbf x}$}. Think, for example, of a 
person shooting $n$ times on a target, and counting, at the end,
the numbers of scores $x$ in order to evaluate his efficiency.
If somebody else fires by mistake at  random on his target,
the number $x$ will be affected by background. 
The same situation can happen in measuring efficiencies in those 
situations (for example due to high rate or 
loose timing) 
in which the time correlation between the equivalents of
`shooting' and `scoring'
cannot be done on a event by event 
basis (think, for example, to neutron or photon detectors).

The problem will be solved assuming that the background is described 
by a Poisson process of well known intensity $r_b$, that corresponds
to a well known expected value $\lambda_b$ of the resulting
Poisson distribution (in the time domain  $\lambda_b=r_b\cdot T$,
where $T$ is measuring time). 
In other words, the observed $x$ is the sum of two contributions:
$x_s$ due to the {\it signal}, binomially distributed with
${\cal B}_{n,p}$, plus $x_b$ due to background, Poisson 
distributed with parameter $\lambda_b$, indicated by ${\cal P}_{\lambda_b}$.

For large numbers (and still relatively low background) 
the problem is easy to solve: we subtract the expected number 
of background and calculate the proportion $\hat p = (x-\lambda_b)/n$.
For small numbers, the `estimator' $\hat p$ can become smaller
than 0 or larger then 1. And, even if  $\hat p$  comes out in the correct
range, it is still affected by large uncertainty. Therefore
we have to go through
a rigorous probability inversion, that in this case is given by
\begin{eqnarray}
 f(p\,|\,n,x,\lambda_b) &\propto&
         f(x=x_s+x_b\,|\,n,p,\lambda_b) \cdot f_\circ(p) \,,
\end{eqnarray}
where we have written explicitly in the likelihood that $x$
is due to the sum of two (individually unobservable!) contributions
$x_s$ and $x_b$
(hereafter the subscripts  $s$ and $b$ stand for
{\it signal} and {\it background}.)

\item[{\it b)}]
{\bf The background can show up, at random, 
as independent `fake' trials, all with the same 
${\mathbf p_b}$ of producing successes}. 
An example, that has indeed prompted this paper, 
is that of the measuring the proportion of blue galaxies
in a small region of sky where there are galaxies belonging to
a cluster, as well as background galaxies, 
the average proportion of blue galaxies of which is well known.
In this case  both $n$ and $x$ have two contributions:
\begin{eqnarray}
n & = & n_s+n_b \\ 
x & = & x_s+x_b 
\end{eqnarray}
with
\begin{eqnarray}
n_b & \sim  & {\cal P}_{\lambda_b} \\ 
x_b & \sim  & {\cal B}n_b,p_b    \\
x_s &  \sim  & {\cal B}n_s,p_s\,,
\end{eqnarray}
where  `$\sim$' stands for `follows a given distribution'.

Again, the trivial large number (and not too large background) 
solution is the proportion
of background subtracted numbers, 
$\hat p = (x-p_b\,\lambda_b)/(n-\lambda_b)$. 
But in the most general case we need to infer $p$ from 
\begin{eqnarray}
 f(p_s\,|\,n,x,\lambda_b,p_b) &\propto&
         f(x=x_s+x_b\,|\,n=n_s+n_b,p_b,\lambda_b) 
\cdot f_\circ(p) \,. \nonumber \\
 && 
\end{eqnarray}
We might  be also interested also to other questions,
like e.g. how many of the $n$ object are due to the signal,
i.e. 
$$f(n_s\,|\,n,x,\lambda_b,p_b)\,.$$ 
Indeed, the general problem
lies in the joint inference 
$$f(n_s,p_s\,|\,n,x,\lambda_b,p_b),$$
from which we can get other information, like the conditional
distribution of $p_s$ for any given number of events 
attributed to signal:  
$$f(p_s\,|\,n,n_s,x,\lambda_b,p_b)\,.$$
Finally, we may also be interested in the rate $r_s$ of the 
signal objects, responsible of the $n_s$  signal objects
in the sample (or, equivalently, to the Poisson distribution
parameter $\lambda_s$):
$$f(\lambda_s\,|\,n,x,\lambda_b,p_b)\,.$$
\end{enumerate}

\section{Inferring $p$ in absence of background}\label{sec:no_bkgd}
\begin{figure}[t]
\centering\epsfig{file=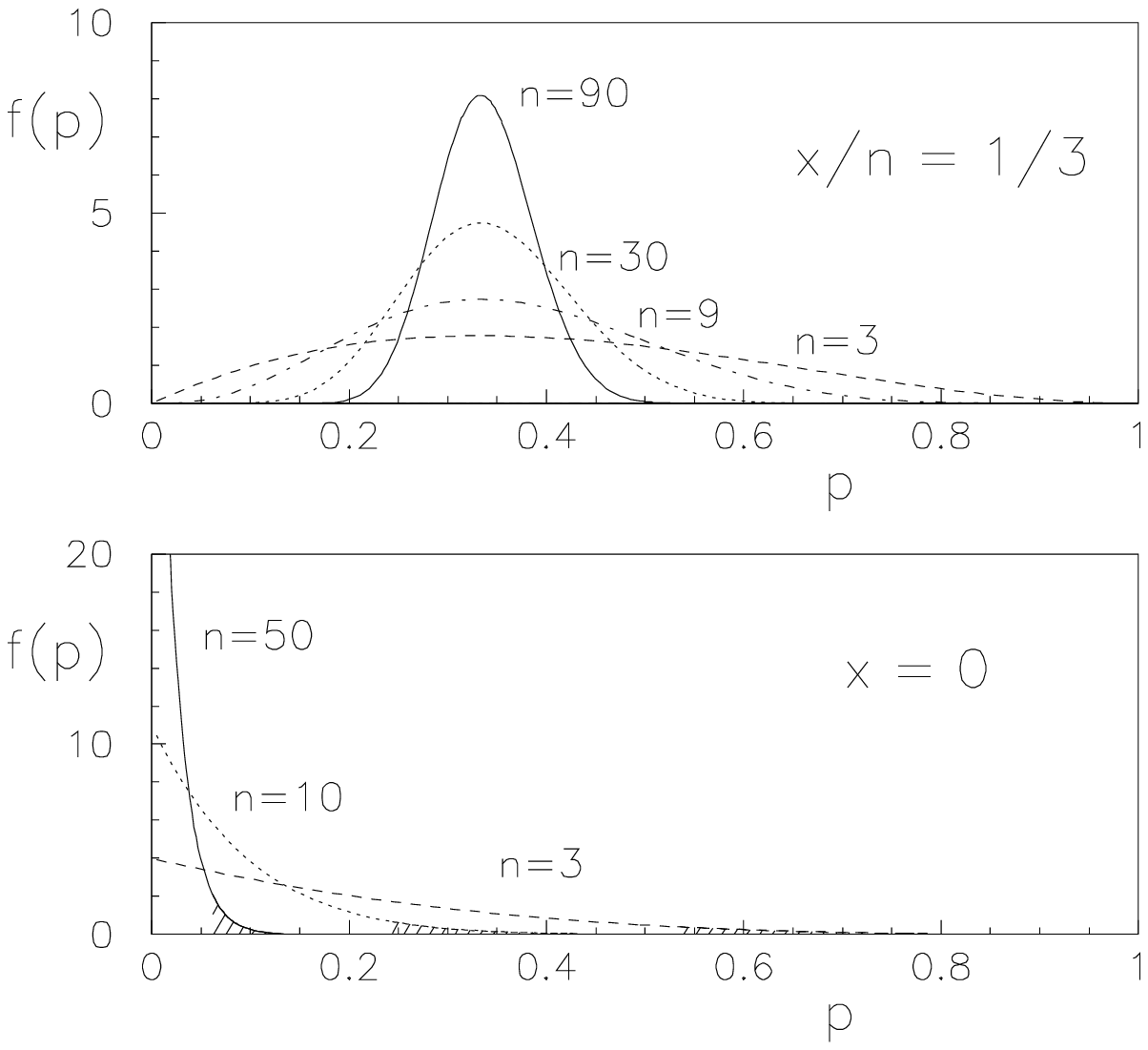,bbllx=13,bblly=161,bburx=356,bbury=317,clip=,width=\linewidth}
\caption{\small Probability density function of the binomial parameter
$p$, having observed $x$ successes in $n$ trials.\cite{BR}}
\label{fig:beta_up}
\end{figure}

The solution of Eq.(\ref{eq:inf_binom}) depends, at least
in principle, on the assumption on the prior $f_\circ(x)$.
Taking a flat prior between 0 and 1, that models our 
{\it indifference} on the possible values of $p$ {\it before}
we take into account the result of the experiment in which
$x$ successes were observed in $n$ trials, we get 
(see e.g. \cite{BR}): 
\begin{equation}
f(p\,|\,x,n,{\cal B}) 
= \frac{(n+1)!}{x!\,(n-x)!}\,p^x\,(1-p)^{n-x}\,,
\label{eq:inv_binom}
\end{equation}
some examples of which are shown in Fig.~\ref{fig:beta_up}.
Expected value, mode (the value of $p$ for which $f(p)$ 
has the maximum) and variance of this distribution are:
\begin{eqnarray}
\mbox{E}(p) &=& \frac{x+1}{n+2} 
\label{eq:infbinom1}\\
\mbox{mode}(p)= p_m & = & x/n \\
\sigma^2(p)=\mbox{Var}(p) &=& \frac{(x+1)(n-x+1)}{(n+3)(n+2)^2} \\
       &=& \mbox{E}(p)\,\left(1 - \mbox{E}(p)\right)\,\frac{1}{n+3}
\label{eq:infbinom2}\,.
\end{eqnarray}
Eq.~(\ref{eq:infbinom1}) 
is known as   
``recursive Laplace formula'',
 or ``Laplace's rule of succession''.
Not that there is no magic if the formula
 gives a sensible result even 
for the extreme cases $x=0$ and $x=n$ for all values of $n$ 
(even if $n=0$\,!).
It is just a consequence of the prior: in absence of new information,
we get out what we put in!

From Fig.~\ref{fig:beta_up} we can see that for large numbers 
(and with $x$ far from 0 and from $n$)
$f(p)$ tends to a Gaussian. This is just the reflex of the limit
to Gaussian of the binomial. In this large numbers limit 
$ \mbox{E}(p) \approx p_m =x/n$ and  $\sigma(p) \approx \sqrt{x/n\,(1-x/n)/n}$. 

\subsection{Meaning and role of the prior: many data limit 
versus frontier type measurements}\label{ss:priors}
One might worry about the role of the prior. Indeed,
in some special cases of importance 
{\it frontier type} measurement one {\it has} to. 
However, in most {\it routine} cases, the prior just plays
the role of a {\it logical tool to allow probability 
inversion}, but it is in fact absorbed in the normalization
constant. (See extensive discussions in Ref.~\cite{BR} 
and references therein.) 

\begin{figure}
\begin{center}
\begin{tabular}{|c|c|}\hline 
& \\
\multicolumn{1}{|l|}{{\bf A)} {\small $r=s=$\,{\bf 1}, 1.1 e 0.9}} & 
\multicolumn{1}{l|}{{\bf B)} {\small $r=s=$\,{\bf 2}, 3, 4, 5}} \\ 
\epsfig{file=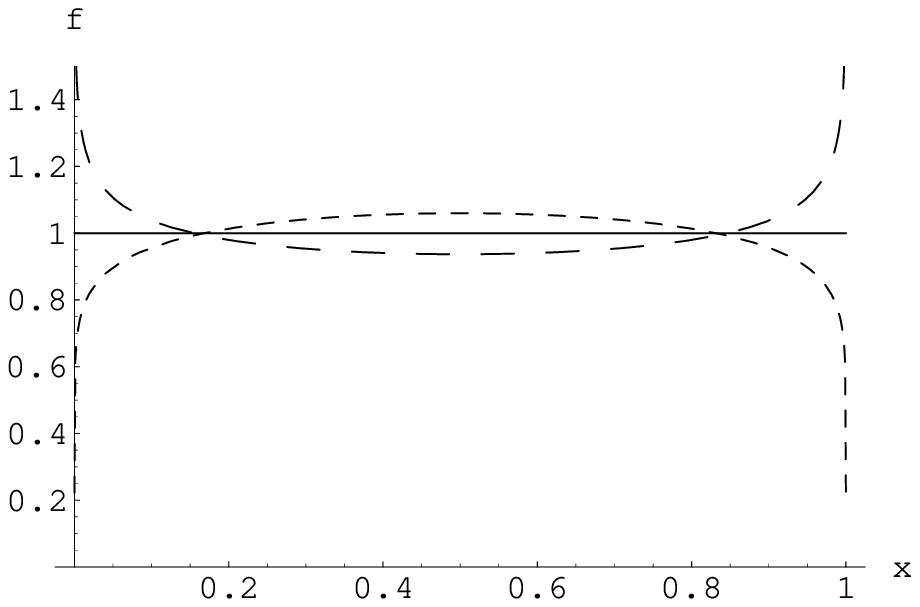,width=0.47\linewidth,clip=} &
\epsfig{file=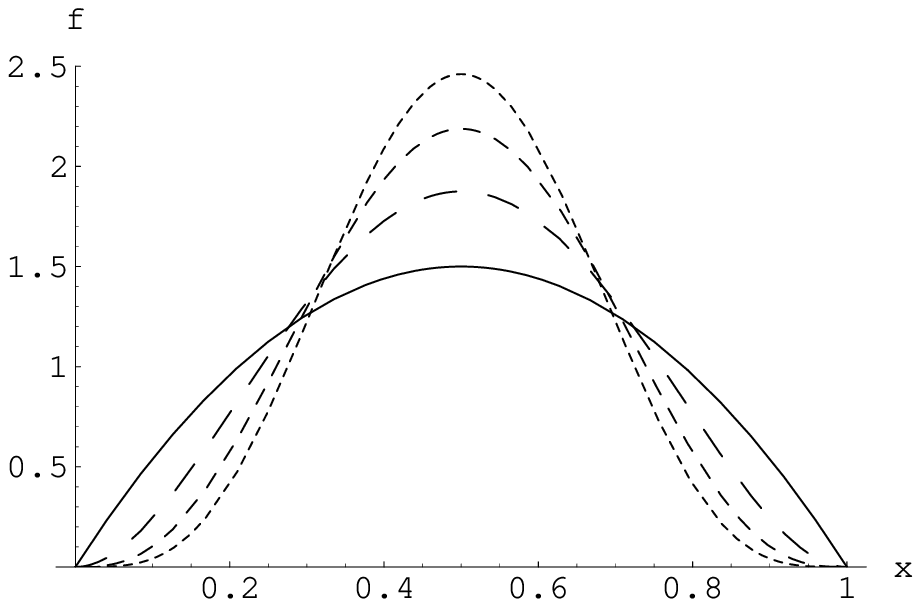,width=0.47\linewidth,clip=} \\ \hline 
& \\
\multicolumn{1}{|l|}{{\bf C)} {\small $r=s=$\,{\bf 0.8}, 0.5, 0.2, 0.1}} & 
\multicolumn{1}{l|}{{\bf D)} {\small $r=0.8$; $s=$\,{\bf 1.2}, 1.5, 2, 3}}\\
\epsfig{file=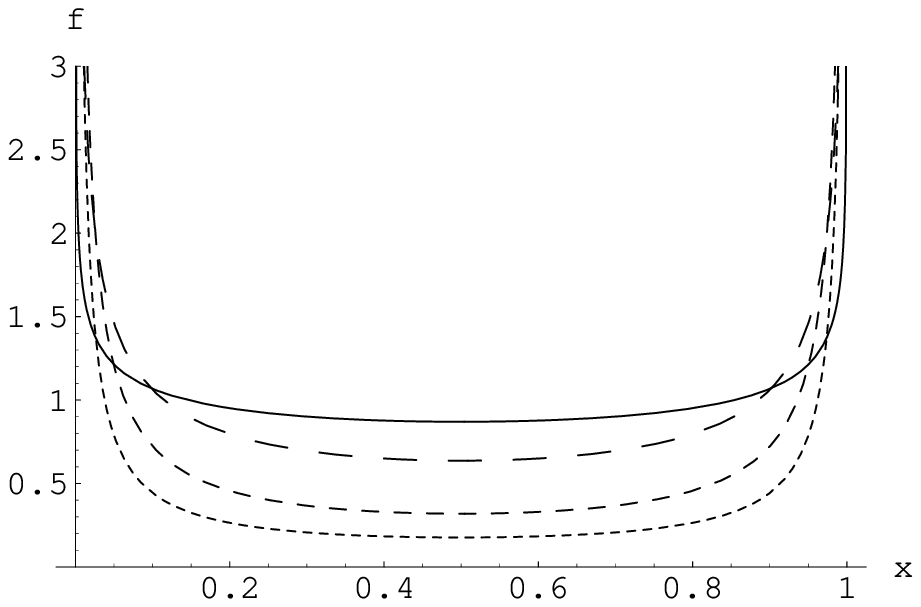,width=0.47\linewidth,clip=} &
\epsfig{file=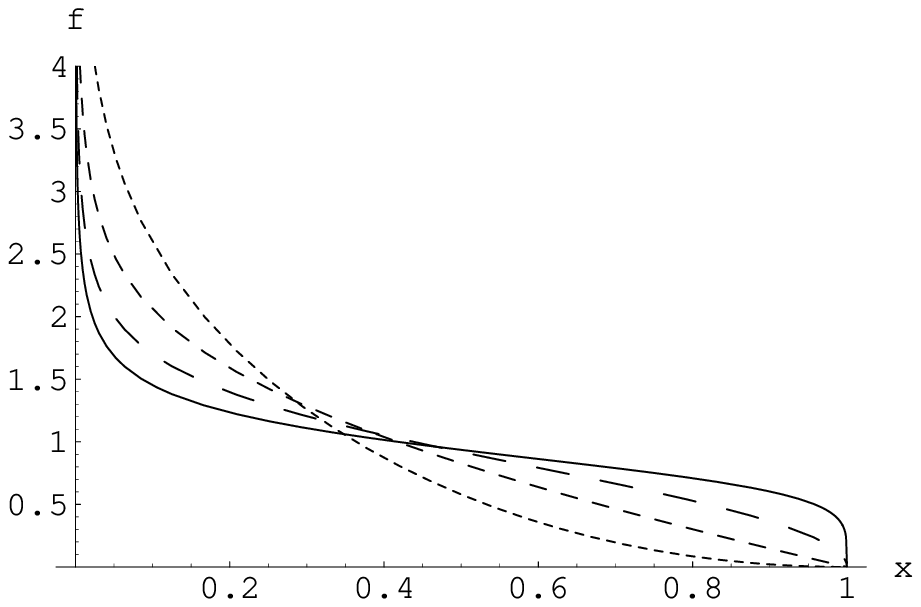,width=0.47\linewidth,clip=} \\ \hline 
& \\
\multicolumn{1}{|l|}{{\bf E)} {\small $(r,\,s)=$ (3,\,5), ({\bf 5,\,5}),
(5,\,3)}} & 
\multicolumn{1}{l|}{{\bf F)} {\small $(r,\,s)=$ (30,\,50), ({\bf 50,\,50}),
(50,\,30)}} \\
\epsfig{file=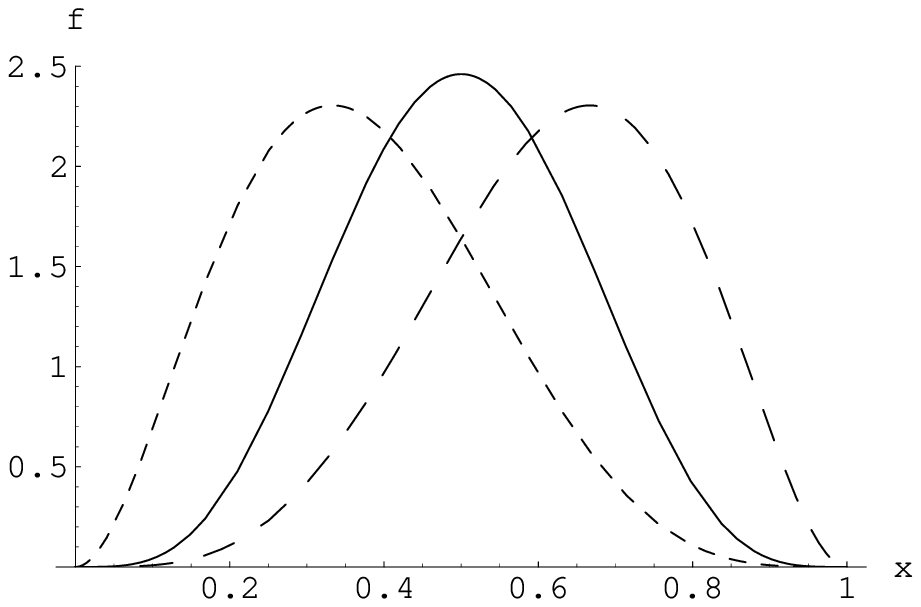,width=0.47\linewidth,clip=} &
\epsfig{file=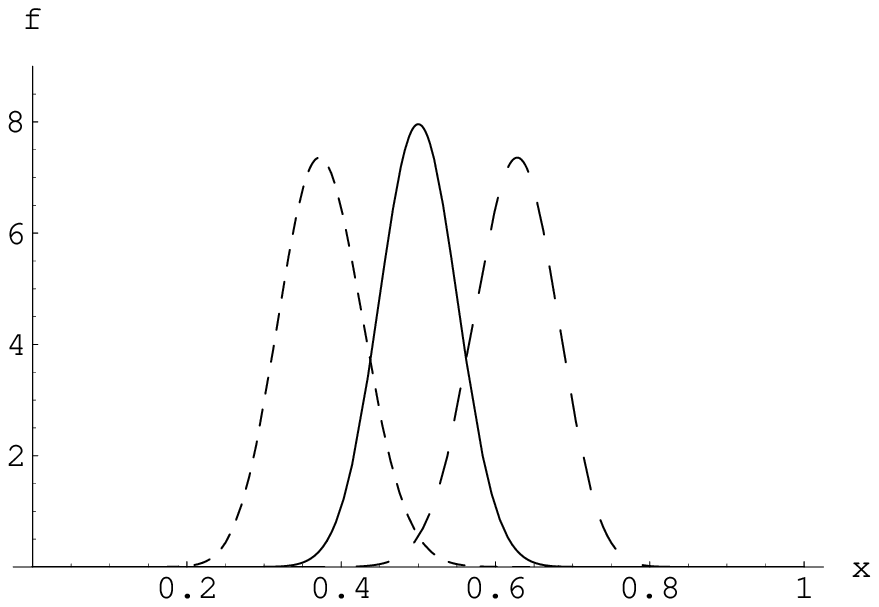,width=0.47\linewidth,clip=} \\ \hline
\end{tabular}
\end{center}
\caption{\small Examples of Beta distributions for some 
values of  $r$ and $s$ \cite{BR}. 
The parameters in bold refer to continuous curves.}
\label{fig:betas}
\end{figure}

In order to see the effect of the prior, let us model it
in a easy and powerful way using a {\it beta} distribution,
a very flexible tool to 
 describe many situations of prior knowledge  
about a variable defined in the interval
between 0 and 1 (see Fig. \ref{fig:betas}). 
The  beta distribution
is the {\it conjugate prior} of the binomial distribution,
i.e. prior and posterior belong to the same function family,
with parameters updated by the data via the likelihood.
In fact, a generic beta distribution in function of the variable $p$
is given by
\begin{equation}
f(p\,|\,\mbox{Beta}(r,s))=\frac{1}{\beta(r,s)}p^{r-1}(1-p)^{s-1}
\hspace{0.6cm}\left\{\!\begin{array}{l}  r,\,s > 0 \\
   0\le p\le 1 \,.  \end{array}\right.
\label{eq:distr_beta}
\end{equation}
The denominator is just for normalization and, indeed, the
integral 
$\beta(r,s)=\int_0^1 p^{r-1}(1-p)^{s-1}\,\mbox{d}p$ defines
the special function beta that names the distribution.
We immediately recognize Eq.~(\ref{eq:inv_binom}) 
as a beta distribution of parameters $r=x+1$ and $s=n-x+1$
[and the fact that $\beta(r,s)$ is equal to
$(r-1)!(s-1)!/(s+r-1)!$ for integer arguments].

For a generic beta we get the following posterior
(neglecting the irrelevant normalization factor):
\begin{eqnarray}
f(p\,|\,n,x,\mbox{Beta}(r,s)) &\propto & 
    \left[p^x (1-p)^{n-x}\right] \times \left[p^{r_i-1}(1-p)^{s_i-1}\right] \\
    &\propto &  p^{x+r_i-1} (1-p)^{n-x+s_i-1}\,,
\end{eqnarray}
where the subscript $i$ stands for {\it initial}, synonym of prior. 
We can then see that the final distribution  
is still a beta with parameters 
$r_f = r_i+x$ and $s_f =s_i+ (n-x) $:
 the first parameter is updated by the 
number of successes, the second parameter by the number 
of failures. 

Expected value, mode and variance of the generic beta 
of parameters $r$ and $s$ are: 
\begin{eqnarray}
\mbox{E}(X)&=&\frac{r}{r+s} \label{eq:Ebeta}\\
\mbox{mode}(X) &=& (r-1)/(r+s-2) \ \ \ \ \ \ [r>1\ \mbox{and}\ s>1] \\
\mbox{Var}(X)&=&\frac{rs}{(r+s+1)\,(r+s)^2}\ \ \ \ \ 
      [r+s > 1] \,.  \label{eq:Varbeta}
\end{eqnarray}
Then we can use these formulae for the beta posterior of parameters
$r_f$ and $s_f$. 

The use of the conjugate prior in this problem 
demonstrates in a clear way
how the inference becomes 
progressively independent from the prior information in the limit of
a large amount of data: 
this happens when both $x\gg r_i$ and $n-x\gg s_i$. In this limit we get
the same result we would get from a flat prior ($r_i=s_i=1$, see
Fig.~\ref{fig:betas}).
For this reason in standard `routine' situation, 
we can quietly and safely take a flat prior. 

\begin{figure}[t]
\centering\epsfig{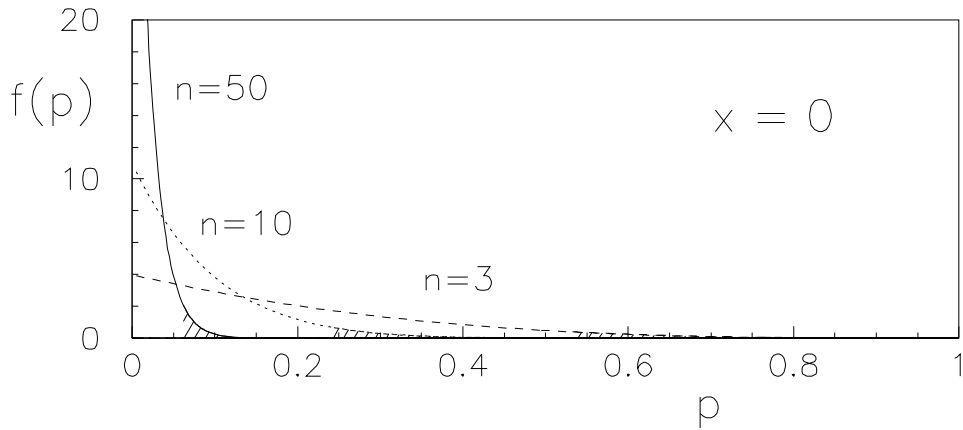}
\caption{\small Probability density function of the binomial parameter
$p$, having observed no successes in $n$ trials.\cite{BR}}
\label{fig:beta_down}
\end{figure}

Instead, the treatment needs much more care in situations
typical of `frontier research': small numbers, and often with
no single `successes'. Let us consider the latter case and let us
assume a na\"\i ve flat prior, that it is considered to
represent `indifference' of the parameter $p$ between 0 and 1.
From Eq.~(\ref{eq:inv_binom}) we get 
\begin{eqnarray}
f(p\,|\,x=0,n,{\cal B},\mbox{Beta}(1,1)) & = & (n+1) \, (1-p)^n \,.
\label{eq:fp_x0}
\end{eqnarray} 
(The prior has been written explicitly among the conditions
of the posterior.) 
Some examples are given in Fig.~(\ref{fig:beta_down}). 
As $n$ increases, $p$ is more and more 
constrained in proximity of 0.
In these cases we are used to give {\it upper limits} at a certain
level of confidence. 
The natural meaning that we give to this
expression is that we are such and such percent confident that 
$p$ is below the reported upper limit. In the Bayesian approach this is
is straightforward, for confidence and probability are synonyms. 
For example, if we want to give the limit that makes us 95\%
sure that $p$ is below it, i.e. $P(p\le p_{u_{0.95}}) = 0.95$,
then we  have to calculate the value $p_{u_{0.95}}$ such that
the cumulative function $F(p_{u_{0.95}})$ is equal to 0.95:
\begin{eqnarray}
 F(p_{u_{0.95}}\,|\,x=0,n,{\cal B},\mbox{Beta}(1,1)) 
&=& \int_0^{p_{u_{0.95}}}f(p)\,\mbox{d}p \\ 
& = &  1 - (1-p_u)^n = 0.95\,,
\end{eqnarray}
that yields 
\begin{eqnarray}
p_{u_{0.95}} & = &  1 - \sqrt[n+1]{0.05}\, .
\end{eqnarray}
For the three examples given in Fig.~\ref{fig:beta_down}, with
$n=3$, 10 and 50, we have $p_{u_{0.95}}=0.53$,
0.24 and 0.057, respectively. 
These results are in order, as long the flat prior reflected our expectations 
about $p$, that it could be about equally likely in any sub-interval
of fixed width  in the interval between
0 and 1 (and, for example, we believe that it is equally
likely below 0.5 and above 0.5).

However, this is often not the case in frontier research. 
Perhaps we were looking for a very rare process, with 
a very small $p$. Therefore, having done only 50 trials, we cannot say
to be 95\% sure that $p$ is below 0.057. In fact, by logic, the previous
statement implies that we are 5\% sure that $p$ is above 0.057, 
and this might seem too much for the scientist expert of the
phenomenology under study. (Never ask mathematicians about priors!
Ask yourselves and the colleagues you believe are the most
knowledgeable experts of what you are studying.) In general I suggest
to make the exercise of calculating a 50\% upper or lower limit,
i.e. the value that divides the possible values in two equiprobable
regions: we are as confident that $p$ is above as it is below 
$p_{u_{0.5}}$. For $n=50$ we have $p_{u_{0.5}}=0.013$. If a physicist 
was looking for a rare process, he/she would be highly
embarrassed to report to be 50\% confident that $p$ is above 0.013.
But he/should be equally embarrassed to report to be 95\% confident
that $p$ is below 0.057, because both statements are logical
consequence of the same result, that is Eq.~(\ref{eq:fp_x0}).
If this is the case, a better grounded prior is needed, instead
of just a `default' uniform. For example one might thing that
several order of magnitudes in the small $p$ range are considered
equally possible. This give rise to a prior that is uniform
in $\ln p$ (within a range $\ln p_{min}$ and  $\ln p_{max}$), 
equivalent to $f_\circ(p)\propto 1/p$
with lower and upper cut-off's.  

\begin{figure}
\centering\epsfig{file=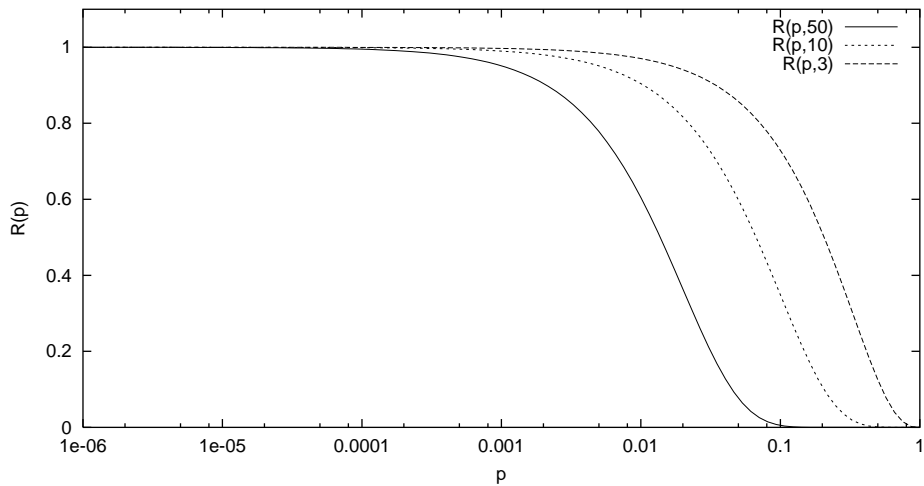,clip=,width=\linewidth}
\caption{\small Rescaled likelihoods for $x=0$ and some values of $n$}
\label{fig:Rp}
\end{figure}

Anyway, instead of playing blindly with mathematics, 
looking around for `objective' priors, or priors that 
come from abstract arguments, it is important to understand at once
the role of prior and likelihood. Priors are logically important
to make a `probably inversion' via the Bayes formula, and 
it is a matter of fact that no other route to probabilistic
inference exists. The task of the likelihood is to modify our beliefs, 
distorting the pdf that models them. 
Let us plot the three
likelihoods of the three cases of Fig.~\ref{fig:beta_down}, 
{\it rescaled} to the asymptotic value $p\rightarrow 0$
(constant factors are irrelevant in likelihoods). 
It is preferable to plot them in a log scale along the
abscissa to remember that several orders of magnitudes are involved
 (Fig.~\ref{fig:Rp}).  

We see from the figure that in the high $p$ region  the beliefs
expressed by the prior are strongly dumped. If we were
convinced that $p$ was in that region we have to 
dramatically review our beliefs. With the increasing 
number of trials, the region of `excluded' values of $\log p$ 
increases too. 
 
Instead, for very small values of $p$,
the likelihood becomes flat, i.e. equal to the asymptotic value
$p\rightarrow 0$. The region of flat likelihood represents the values of $p$ 
for which the experiment loses sensitivity: if 
scientific motivated priors concentrate the probability
mass in that region, then the experiment is irrelevant 
to change our convictions about $p$.

Formally the rescaled likelihood
\begin{eqnarray}
{\cal R}(p;\, n,\,x=0) & = & \frac{f(x=0\,|\,n,\,p)}
                                  {f(x=0\,|\,n,\,p\rightarrow 0)}\,,
\label{eq:R_bin}
\end{eqnarray}
equal to $(1-p)^n$ in this case,
is a functions that gives the Bayes factor of a generic $p$ with respect
to the reference point $p=0$ for which the experimental
sensitivity is certainly lost. Using the Bayes formula, 
${\cal R}(p;\, n,\,x=0)$
 can rewritten as
\begin{eqnarray}
{\cal R}(p;\, n,\,x=0) & = & \frac{f(p\,|\,n,\,x=0)}
                                  {f_\circ(p)} \left/
                             \frac{f(p=0\,|\,n,\,x=0)}
                                  {f_\circ(p=0)} \right.\,,
\end{eqnarray}
to show that it can be interpreted as a {\it relative belief
updating factor}, in the sense that it gives the updating
factor for each value of $p$ with respect to that 
at the asymptotic value $p\rightarrow 0$. 

We see that this ${\cal R}$ function gives a way to report
an upper limit that do not depend on prior: it can be any conventional
value in the region of transition from ${\cal R}=1$ to 
${\cal R}=0$. However, this limit cannot have a probabilistic 
meaning, because does not depend on prior. It is instead a 
{\it sensitivity bound}, roughly separating the excluded
high $p$ value from the the small $p$ values about which the 
experiment has nothing to 
say.\footnote{``{\it Wovon man nicht reden kann, 
dar\"uber muss man schweigen}'' (L. Wittgenstein).}

For further discussion about the role of prior in 
frontier research, applied to the Poisson process, see
Ref.~\cite{RPoisson}. For examples of experimental
results provided with the ${\cal R}$ function, 
see Refs.~\cite{Zeus_ci,Higgs,Beppo}.

\section{Poisson background on the observed number of 
 `successes'}\label{sec:bkgd_x}
Imagine now that the $x$ successes might contains an unknown number
of background events $x_b$, of which  we only know their
expected value $\lambda_b$, estimated somehow and 
about which we are quite sure (i.e. uncertainty about $\lambda_b$
is initially neglected --- it will be indicated at the end 
of the section how to handle it).
We make the assumption that the background events
come at random and are described by a Poisson process of
intensity $r_b$, such that the Poisson parameter 
$\lambda_b$ is equal to 
$r_b\times \Delta T$
in the domain of time, with $\Delta T$ the 
observation time. (But we could as well reason in 
other domains, like objects per unit of length, surface, 
volume, or solid angle. The density/intensity parameter $r$ 
will have different dimensions depending on the context, while 
$\lambda$ will always be dimensionless.)
 
The number of observed successes $x$ has now two contributions:
\begin{eqnarray}
x &=& x_s + x_b \\
x_s & \sim & {\cal B}_{n,p} \\
x_b & \sim & {\cal P}_{\lambda_b}\,,
\end{eqnarray}
In order to use Bayes theorem we need to calculate 
$f(x\,|\,n,\,p,\,\lambda_b)$, that is
$f(x= x_s + x_b\,|\,{\cal B}_{n,p},\,{\cal P}_{\lambda_b})$, 
i.e. is the probability function of the sum 
of a binomial variable and a Poisson variable.
The combined probability function is give by (see e.g. section 4.4
of Ref.~\cite{BR}):
\begin{eqnarray} 
f(x\,|\,{\cal B}_{n,p},\,{\cal P}_{\lambda_b}) &=& 
\sum_{x_s,\,x_b} \delta_{x,\,x_s+x_b} \,f(x_s\,|\,{\cal B}_{n,p_s})
\,f(x_b\,|\,{\cal P}_{\lambda_b})
\label{eq:bin+pois}
\end{eqnarray}
where $\delta_{x,\,x_s+x_b}$ is the Kronecker delta that constrains
the possible values of 
$x_s$ and $x_b$ in the sum ($x_s$ and $x_b$ run from 0 to 
the maximum allowed by the constrain).
Note that we do not need to 
calculate this probability function for all $x$, but only 
for the number of actually observed successes. 

\begin{figure}
\begin{center}
\begin{tabular}{c}
\epsfig{file=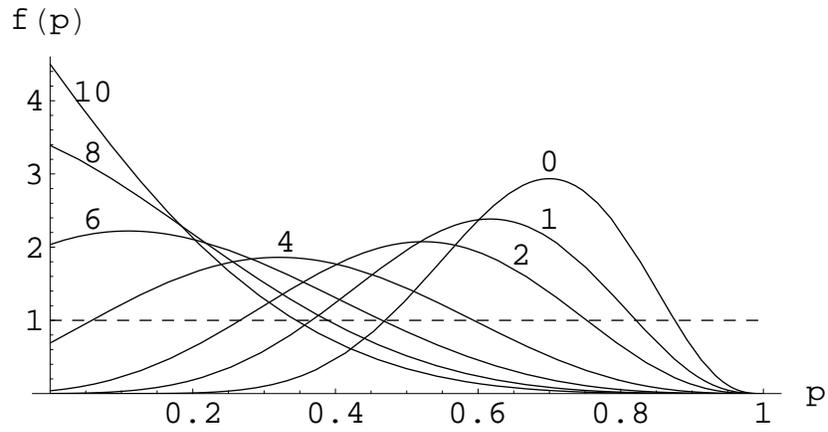,clip=,width=\linewidth}\\
\mbox{} \\
\mbox{} \\
\epsfig{file=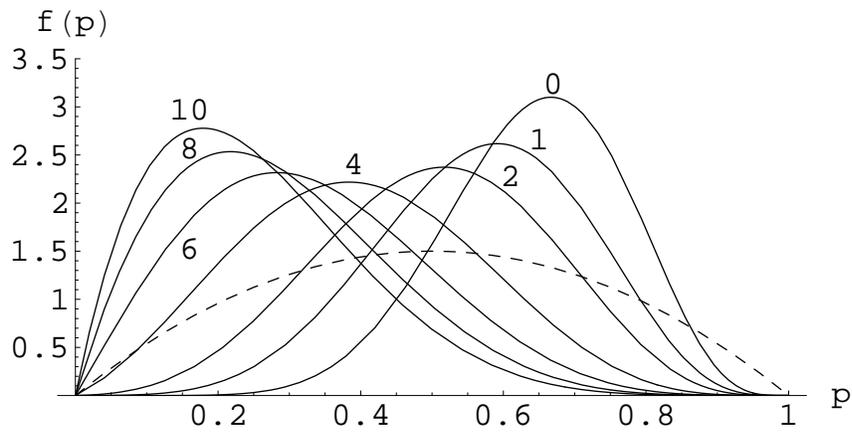,clip=,width=\linewidth}
\end{tabular}
\end{center}
\caption{\small Inference of $p$ for $n=10$, $x=7$,  
and several hypotheses of background
(right to left curves for $\lambda_B=0,\,1,\,2,\,4,\,5,\,6,\,10$)
and two different priors (dashed lines), $\mbox{Beta}(1,1)$
in the upper plot and $\mbox{Beta}(2,2)$ in the lower plot
(see text).}
\label{fig:bin_back}
\end{figure}

The inferential result about $p$ is finally given by
\begin{eqnarray} 
f(p\,|\,n,\,p,\,\lambda_b) & \propto & 
f(x\,|\,{\cal B}_{n,p},\,{\cal P}_{\lambda_b})\,f_0(p)\,.
\end{eqnarray}
An example is shown in Fig.~\ref{fig:bin_back}, for $n=10$, $x=7$ and an
expected number of background events 
ranging between 0 and 10, as described in the
figure caption. 
The upper plot of the figure is obtained by a uniform prior
(priors are represented with dashed lines in this figure). 
As an exercise, let us also show in the lower plot of the figure
the results obtained using a broad prior still centered
at $p=0.5$, but that excludes the extreme values 0 and 1, 
as it is often the case in practical cases. 
This kind of prior has been modeled here with a beta function 
of parameters $r_i=2$ and $s_i=2$. 

For the cases of expected background different from zero
 we have also evaluated 
the ${\cal R}$ function, defined in 
analogy to Eq.~(\ref{eq:R_bin}) as 
${\cal R}(p;\, n,\,x,\,\lambda_b) =
 f(x\,|\,n,\,p,\lambda_b)/f(x\,|\,n,\,p\rightarrow 0,\,\lambda_b)\,.$
Note that, while  Eq.~(\ref{eq:R_bin}) is only defined 
for $x\ne 0$, since a single observation makes $p=0$ impossible, 
that limitation does not hold any longer 
in the case of not null expected background.
In fact, it is important
to remember that, as soon as we have background, 
there is some chance
that all observed events are due to it
(remember that a Poisson variable is defined for all non negative
integers!).
This is essentially the reasons why in this case the 
likelihoods tend to a positive value for $p\rightarrow 0$
(I like to call `open' this kind of likelihoods~\cite{BR}). 
\begin{figure}
\begin{center}
\epsfig{file=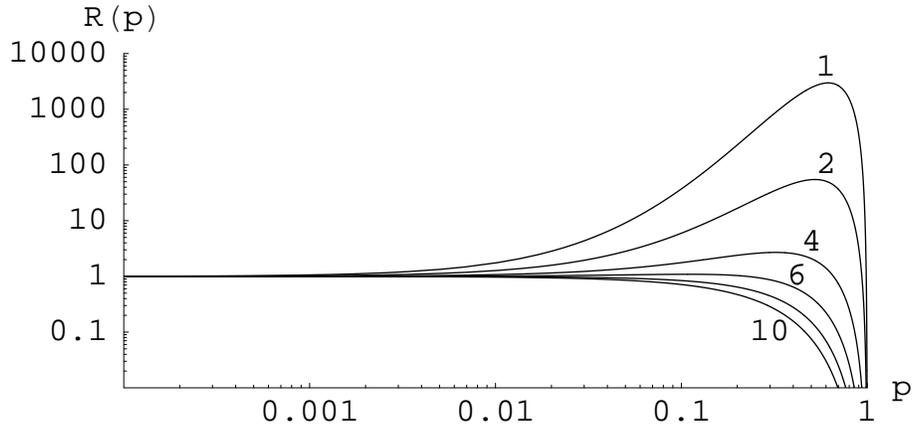,clip=,width=\linewidth}
\end{center}
\caption{\small Relative believe updating factor
 of $p$ for $n=10$, $x=7$ and 
several hypotheses of background:
$\lambda_B=1,\,2,\,4,\,6,\,8,\,10$.}
\label{fig:bin_back_R}
\end{figure}
As discussed above, the power of the data to update the believes on $p$
is self-evident in a log-plot. We seen in Fig. \ref{fig:bin_back_R} that,
essentially, the data do not provide any  
relevant information for values of $p$ below 0.01.

Let us also see what happens when the prior concentrates our beliefs at small
values of $p$, though in principle allowing all values of from 0 to 1. 
Such a prior can be modeled with a log-normal distribution of suitable
parameters (-4 and 1), i.e. 
$f_0(p) = \exp\left[-(\log{p}+4)^2)/2\right]/(\sqrt{2\,\pi}\,p)$,
with an upper cut-off at $p=1$ (the probability that such 
a distribution gives a value above 1 is $3.2\,10^{-5}$). 
Expected value and standard deviation of Lognormal(-4,1) are
0.03 and 0.04, respectively. 
\begin{figure}
\centering\epsfig{file=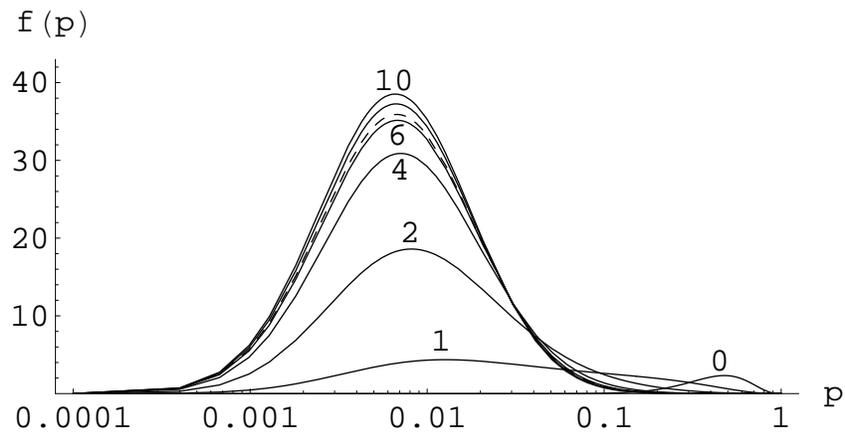,clip=,width=\linewidth}
\caption{\small Inference of $p$ for $n=10$, $x=7$, 
assuming a log-normal prior (dashed line) peaked at low $p$, and with
 several hypotheses of background
($\lambda_B=0, 1,\,2,\,4,\,6,\,8,\,10$).}
\label{fig:bin_back_lognorm_prior}
\end{figure}
The result is given in Fig.~\ref{fig:bin_back_lognorm_prior}, 
where the prior is indicated with a dashed line.

We see that, with increasing expected background, the posteriors are 
essentially equal to the prior. Instead, in case of null background,
ten trials are already sufficiently to dramatically change our prior 
beliefs. For example, initially there was  4.5\% probability
that $p$ was above 0.1. Finally there is only 0.09\% probability 
for $p$ to be below 0.1. 

\begin{figure}
\centering\epsfig{file=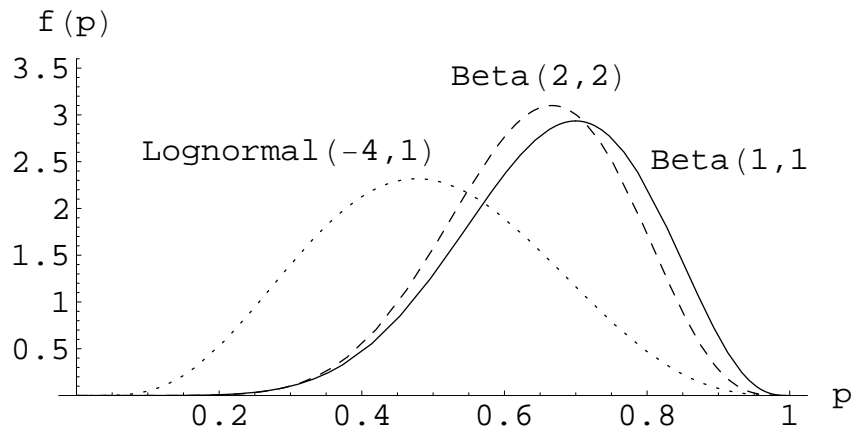,clip=,width=\linewidth}
\caption{\small Inference of $p$ for $n=10$, $x=7$ in absence of background,
with three different priors.}
\label{fig:bin_no_bk_3priors}
\end{figure}

The case of null background is also shown in 
Fig.~\ref{fig:bin_no_bk_3priors}, where the results of the 
three different priors are compared. 
We see that passing from a $\mbox{Beta}(1,1)$ to a $\mbox{Beta}(2,2)$, 
makes little change in the conclusion. Instead, a log-normal prior
distribution peaked at low values of $p$ changes quite a lot the shape
of the distribution, but not really the substance of the result
(expected value and standard deviation 
for the three cases are: 0.67, 0.13; 0.64, 0.12; 0.49, 0.16). 
Anyway, the prior does correctly its job and there should be
no wonder that the final pdf drifts somehow to the left side, 
to take into account a prior knowledge according to
 which 7 successes in 
10 trials was really a `surprising event'. 

\begin{figure}
\centering\epsfig{file=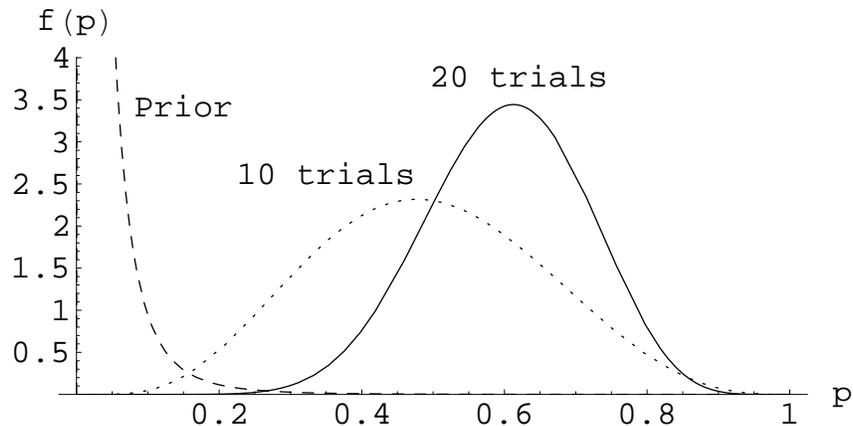,clip=,width=\linewidth}
\caption{\small Sequential inference of 
$p$, starting from a prior peaked at low values, 
given two experiments, each with $n=10$ and  $x=7$.}
\label{fig:bin_no_bk_sequential}
\end{figure}

Those who share such a prior need more solid data to be convinced that 
$p$ could be much larger than what they initially believed. 
Let make the exercise of looking at what happens if a second
experiment gives exactly the same outcome ($x=7$ with $n=10$).
The Bayes formula is applied sequentially, i.e. the posterior
of the first inference become the prior of the second inference. 
That is equivalent to multiply the two priors (we assume 
conditional independence of the two observations).  
The results are given in Fig.~\ref{fig:bin_no_bk_sequential}.
(By the way, the final result is equivalent to having observed 
14 successes in 20 trials, as it should be --- the correct updating
property is one of the intrinsic nice features of the Bayesian approach).

\subsection{Uncertainty on the expected background}
In these examples we made the assumption that the expected number
of background events is well known. If this is not the case, 
we can quantify our uncertainty about it by a pdf $f(\lambda_b)$,
whose modeling depends on our best knowledge about $\lambda_s$.
Taking account of this uncertainty in a probabilistic approach
is rather simple, at least conceptually (calculations can
be quite complicate, but this is a different question). 
In fact, applying  probability theory we get:
\begin{eqnarray}
f(p\,|\,x,\,n) &=& 
\int_0^\infty\!\! f(p\,|\,x,\,n,\,\lambda_b)\,f(\lambda_b)
                  \,\mbox{d}\lambda_b\,.
\end{eqnarray}
We recognize in this formula that the pdf that takes 
into account all possible values of $\lambda$
is a weighted average of all $\lambda_b$ dependent pdf's,
with a weight equal to $f(\lambda_b)$. 

\section{Poisson background on the observed number of `trials' 
and of `successes'}\label{sec:bkgd_n_x}
Let us know move to problem {\it b)} of the introduction. 
Again, we consider only the background parameters are well
known, and refer to the previous subsection for treating
their uncertainty.
To summarize, that is what we assume to know with certainty:
\begin{description}
\item[$n$]: the total observed numbers of `objects', $n_s$ of which
are due to  signal and $n_b$ to background; but these
two numbers are not directly observable and can only be inferred;
\item[$x$]: the total observed numbers of the `objects' of the
subclass of interest, sum of the unobservable $x_s$ and $x_b$;
\item[$\lambda_b$]: the expected number of background objects;
\item[$p_b$]: the expected proportion of successes
due to the background events.
\end{description}
As we discussed in the introduction, we are interested in inferring
the number of signal objects $n_s$, as well as 
the parameter $p_s$ of the `signal'.
We need then to build a likelihood that connects the observed
numbers to all 
quantities we want to infer. Therefore we need to calculate the 
probability function 
$f(x\,|\,n,\ n_s,\, p_s,\, \lambda_b,\, p_b)$. 

Let us first calculate the probability function 
$f(x\,|\,n_s,\,p_s\,n_b,\,p_b)$
that depends on the unobservable $n_s$ and $n_b$.
This is the probability function 
of the sum of two binomial variables:
\begin{eqnarray}
\hspace{-5mm}f_{2{\cal B}}(x\,|\,n_s,\,p_s\,n_b,\,p_b) 
\! &=& \!\sum_{x_s,\,x_b}
\delta_{x,\,x_s+x_b}\,
f(x_s\,|\,{\cal B}_{n_s,\,p_s})
\cdot f(x_b\,|\,{\cal B}_{n_b,\,p_b})\,,
\label{eq:bin+bin}
\end{eqnarray}
where $x_s$ ranges between $0$ and $n_s$, 
and  $x_b$ ranges between $0$ and $n_b$.
$x$ can vary between 0 and $n_s+n_b$, has expected value 
$\mbox{E}(x)=n_s\,p_s+n_b\,p_b$ and variance 
$\mbox{Var}(x) =  n_s\,p_s\,(1-p_s)+n_b\,p_b\,(1-p_b)$.
As for Eq.~(\ref{eq:bin+pois}), 
we need to evaluate
Eq.~(\ref{eq:bin+bin}) only for the observed number of successes.
Contrary to the implicit convention within this paper
to use the same symbol $f(\cdot)$ meaning different 
probability functions and pdf's, we name Eq.~(\ref{eq:bin+bin})
$f_{2{\cal B}}$
for later convenience.

In order to obtain the general likelihood we need, two observations
are in order:
\begin{itemize}
\item
Since $x$ depends from $\lambda$ only via $n_b$, 
then 
$f(x\,|\,n_s,\,p_s\,n_b,\,p_b,\,\lambda_b)$
is equal to $f_{2{\cal B}}(x\,|\,n_s,\,p_s\,n_b,\,p_b)$.
\item
The likelihood that depends also on  $n$ can obtained from \\ 
$f(x\,|\,n_s,\,p_s\,n_b,\,p_b,\,\lambda_b)$ by the following reasoning:
 \begin{itemize} 
  \item
  if $n=n_s+n_b$, then 
   $$f(x\,|\,n,\,n_s,\,p_s\,n_b,\,p_b,\,\lambda_b) = f(x\,|\,n_s,\,p_s\,n_b,\,p_b,\,\lambda_b)\,;$$
  \item
  else $$f(x\,|\,n,\,n_s,\,p_s\,n_b,\,p_b,\,\lambda_b) = 0\,.$$
  \end{itemize}
\end{itemize}
 It follows that
\begin{eqnarray}
\!\!\!f(x\,|\,n,\,n_s,\,p_s,\,n_b,\,p_b,\,\lambda_b) &=& 
f(x\,|\,n_s,\,p_s\,n_b,\,p_b,\,\lambda_b) \,\delta_{n,\,n_s+n_b}\\
&=& f_{2{\cal B}}(x\,|\,n_s,\,p_s\,n_b,\,p_b) \,\delta_{n,\,n_s+n_b}\,.
\end{eqnarray}

At this point we get rid of $n_b$ in the conditions, taking 
account its possible values and their probabilities, given $\lambda_b$:
\begin{eqnarray}
\hspace{-13mm}f(x\,|\,n,\,n_s,\,p_s,\,p_b,\,\lambda_b) &=& 
\sum_{n_b} f(x\,|\,n,\,n_s,\,p_s,\,n_b,\,p_b,\,\lambda_b) \, 
 f(n_b\,|\,{\cal P}_{\lambda_b}) \,, \label{eq_lik_x_5var}
\end{eqnarray}
i.e. 
\begin{eqnarray}
\hspace{-7mm}f(x\,|\,n,\,n_s,\,p_s,\,p_b,\,\lambda_b) &=& \sum_{n_b} f_{2{\cal B}}(x\,|\,n_s,\,p_s\,n_b,\,p_b) 
\, f(n_b\,|\,{\cal P}_{\lambda_b})\,\delta_{n,\,n_s+n_b}\,,
\label{eq_lik_x_5var_e}
\end{eqnarray}
where $n_b$ ranges between 0 and $x$, due to the 
$\delta_{n,\,n_s+n_b}$ condition.
Finally, we can use Eq.~(\ref{eq_lik_x_5var_e}) in Bayes theorem
to infer $n_s$ and $p_s$:
\begin{eqnarray}
f(n_s,\,p_s\,|\,x,\,n,\,\lambda_b,\,p_b)  
&\propto& f(x\,|\,n,\,n_s,\,p_s,\,p_b,\,\lambda_b) \,
  f_0(n_s,\,p_s)
\label{eq:infer_ns_ps}  \\
f(p_s\,|\,x,\,n,\,\lambda_b,\,p_b) &=& 
 \sum_{n_s}f(n_s,\,p_s\,|\,x,\,n,\,\lambda_b,\,p_b)
\label{eq:infer_ps}  \\
f(n_s\,|\,x,\,n,\,\lambda_b,\,p_b) &=&
\int f(n_s,\,p_s\,|\,x,\,n,\,\lambda_b,\,p_b) \, \mbox{d}p_s
\label{eq:infer_ns}   \\
f(p_s\,|\,x,\,n,\,n_s,\,\lambda_b,\,p_b) &=& 
\frac{f(n_s,\,p_s\,|\,x,\,n,\,\lambda_b,\,p_b)}
     {f(n_s\,|\,x,\,n,\,\lambda_b,\,p_b)}\,.
 \label{eq:infer_ps_given_ns}
\end{eqnarray}
We give now some numerical examples. For simplicity (and because
we are not thinking to a specific physical case) we take 
uniform priors, i.e. 
 $f_0(n_s,\,p_s) = \mbox{\it const}$. 
We refer to section \ref{ss:priors} for an extensive discussion
on prior and on critical `frontier' cases.

\subsection{Inferring $p_s$}
\begin{figure}
\begin{center}
\begin{tabular}{c}
\epsfig{file=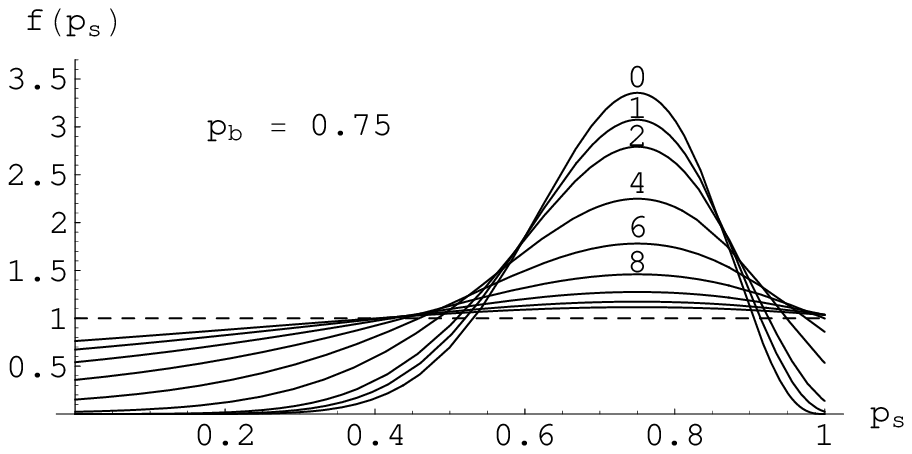,clip=,width=0.975\linewidth}\\
\epsfig{file=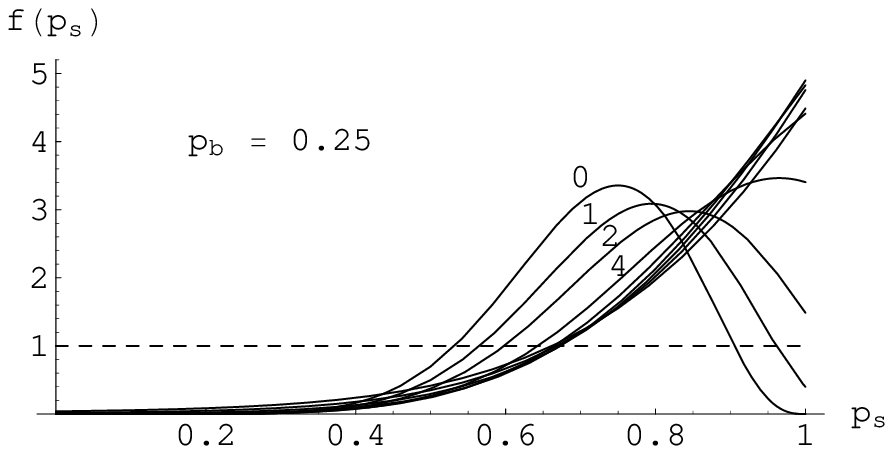,clip=,width=0.975\linewidth}\\
\epsfig{file=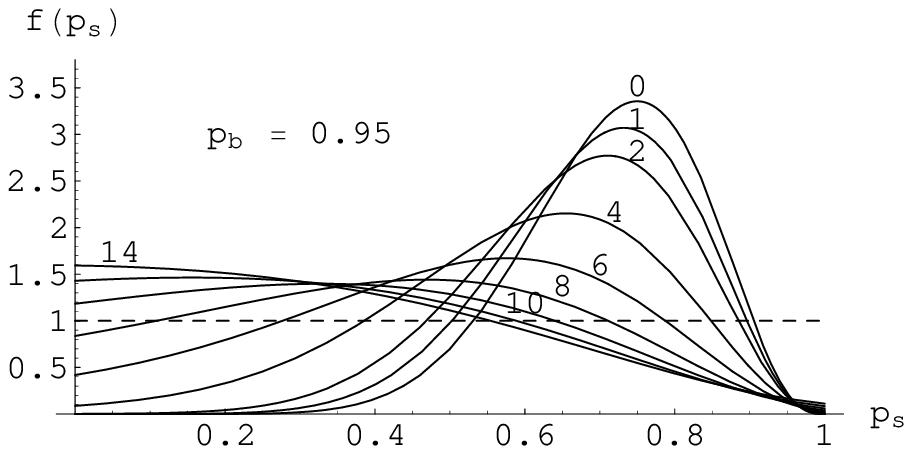,clip=,width=0.975\linewidth} 
\end{tabular}
\end{center}
\caption{\small Inference about $p_s$ for $n=12$ and $x=9$, depending 
on the expected background [$\lambda_b=0$, 1, 2, 4, 6, 8, 10, 14, 
as (possibly) 
indicated by the number above the lines]. The three plots are obtained
by three different hypotheses of $p_b$.}
\label{fig:bin_back_pb075_025_095}
\end{figure}

If priors are uniform then, Eq.~(\ref{eq:infer_ps}) becomes
\begin{eqnarray}
\hspace{-6.0mm}f(p_s\,|\,x,\,n,\,\lambda_,\,p_b) &\propto& 
\sum_{n_s,\,n_b} f_{2{\cal B}}(x\,|\,n_s,\,p_s\,n_b,\,p_b) 
\, f(n_b\,|\,{\cal P}_{\lambda_b})\,\delta_{n,\,n_s+n_b}\,.
\end{eqnarray}
Figure \ref{fig:bin_back_pb075_025_095} gives the result
for  $x=9$, $n=12$, and assuming several hypothesis for $\lambda_b$
and $p_b$.  

\begin{itemize}
\item
The upper plot is for $p_b=0.75$, equal to $x/n$. 
The curves are for $\lambda_b=0,$ 1, 2, 4, 6, 8, 10, 12
and 14, with the order indicated (whenever possible)
in the figure. 
If the expected background is null, we recover the
simple result we already know. As the expected background
increases, $f(p_s)$ gets broader, because the inference
is based on a smaller number of objects attributed
to the signals and because we are uncertain on the
number of events actually due the background. 
In a very noisy environments ($\lambda_b \approx n$, 
or even larger), the data provide very little information
about $p_s$ and, essentially, the prior pdf 
(dashed curve) is recovered. Note also that for
all values of $\lambda_b$ the posterior $f(p_s)$
is peaked at $x/n=0.75$. This is due to the 
fact that $p_b$ was equal to the observed ratio $x/n$,
therefore, for any hypothesis of $n_b$ attributed
to the background, $x_b=p_b \,n_b$ 
counts are in average `subtracted' from $x$ 
(this is properly done in an automatic way 
in the Bayes formula, followed by
marginalization). 
\item
The situation gets more interesting
when $p_b$ differs from $x/n$.

The middle plot in the figure is for $p_b=0.25$.  
Again, the case  $\lambda_b=0$ gives the 
the pdf we already know. But as soon as some
background is hypothesized, the curves start
to drift to the right side. That is because
high background with low $p_b$ favors large
values of $p_s$. 

The opposite happens if we think that background
is characterized by large $p_b$, as shown in
the bottom plot of the figure.
\end{itemize}

\subsection{Inferring $n_s$ and $\lambda_s$}
\begin{figure}
\begin{center}
\begin{tabular}{|c|c|}
\hline
\epsfig{file=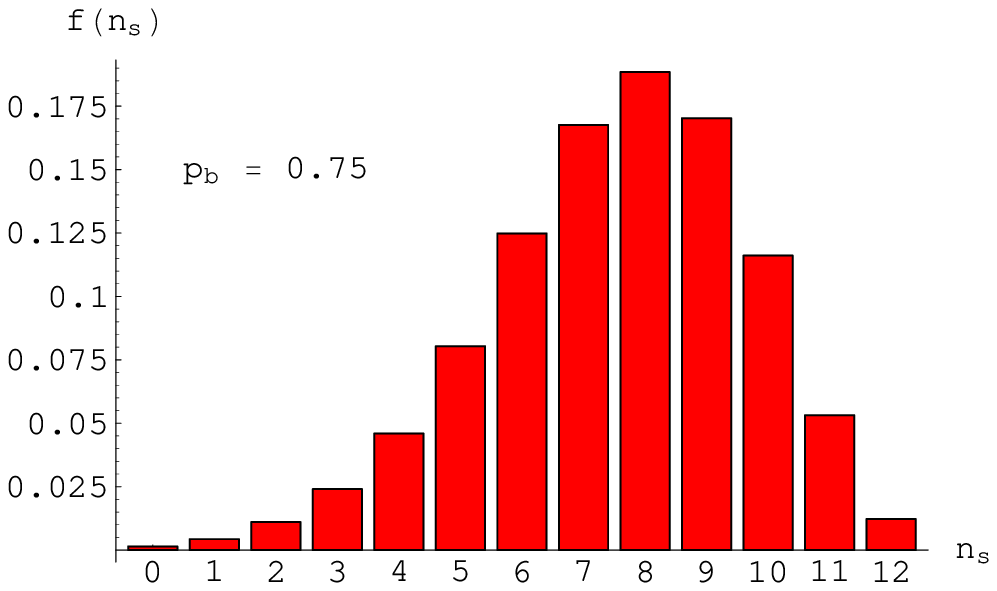,clip=,width=0.46\linewidth} &
\epsfig{file=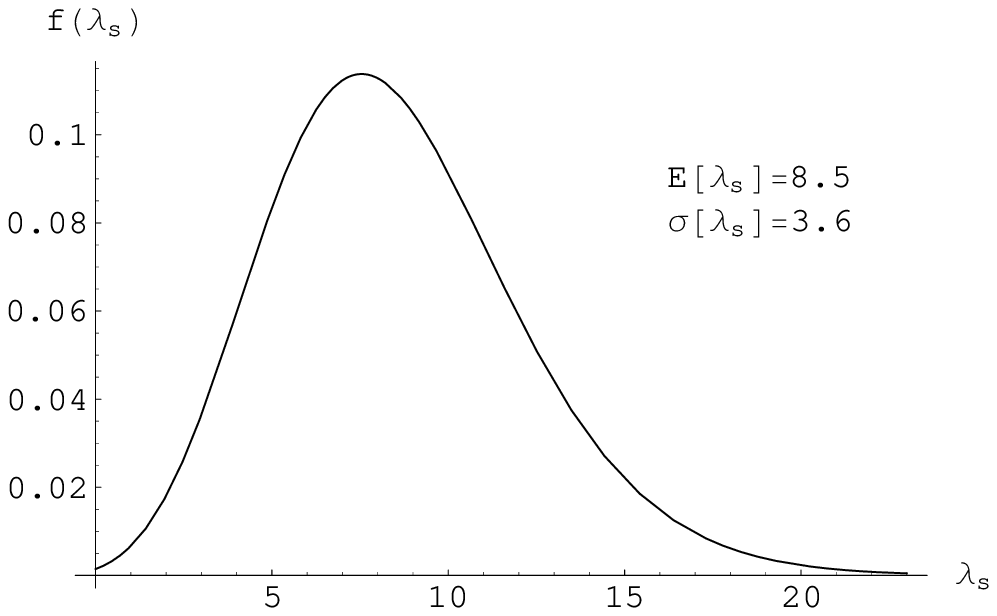,clip=,width=0.46\linewidth} \\
\hline
\epsfig{file=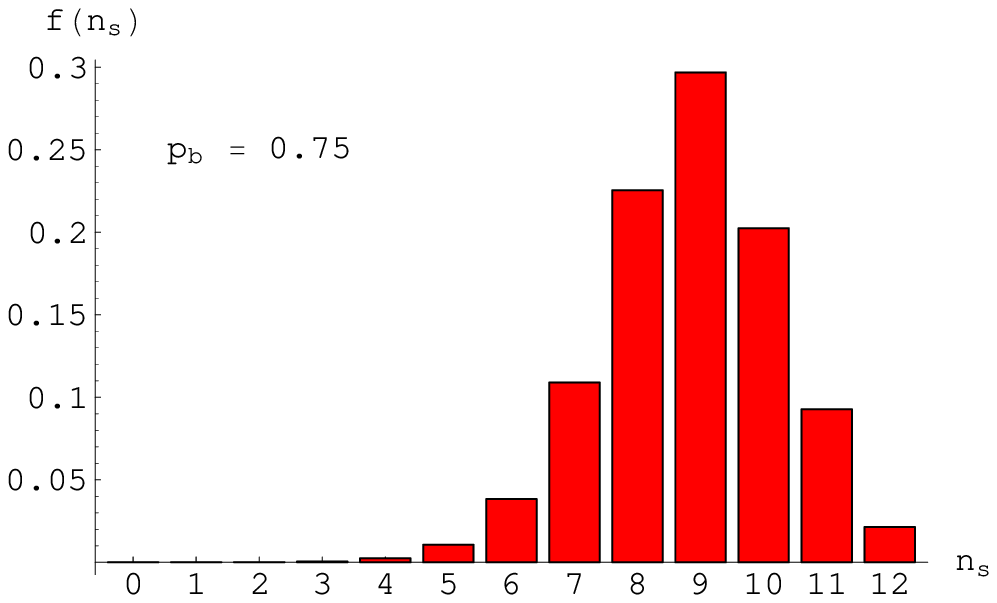,clip=,width=0.46\linewidth} &
\epsfig{file=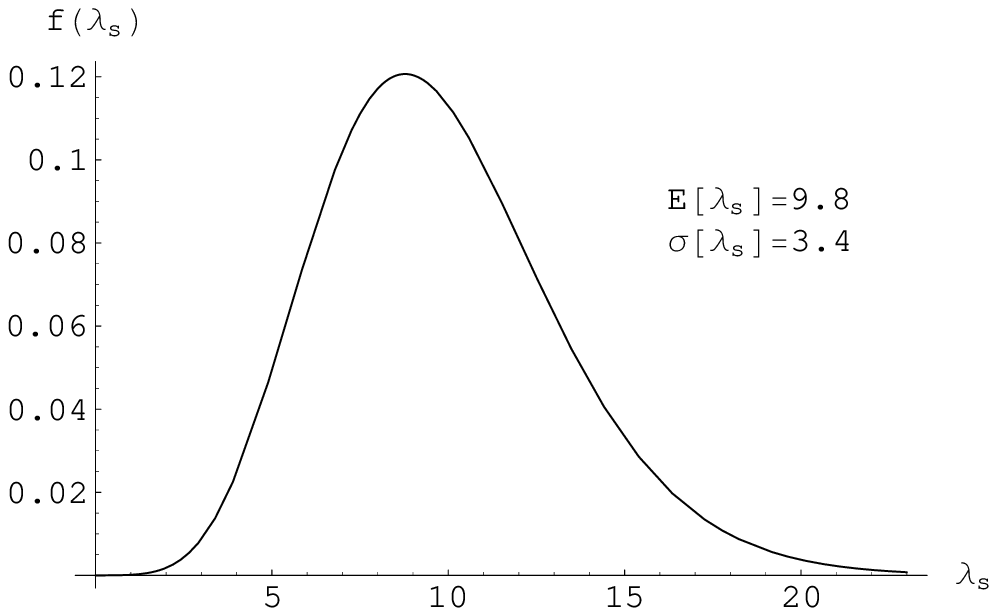,clip=,width=0.46\linewidth} \\
\hline
\epsfig{file=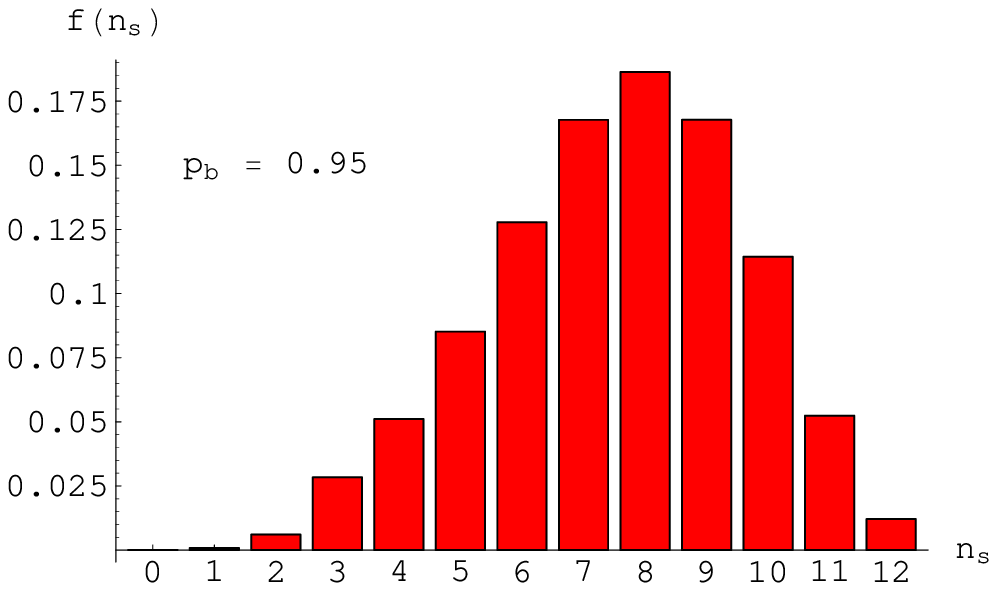,clip=,width=0.46\linewidth} &
\epsfig{file=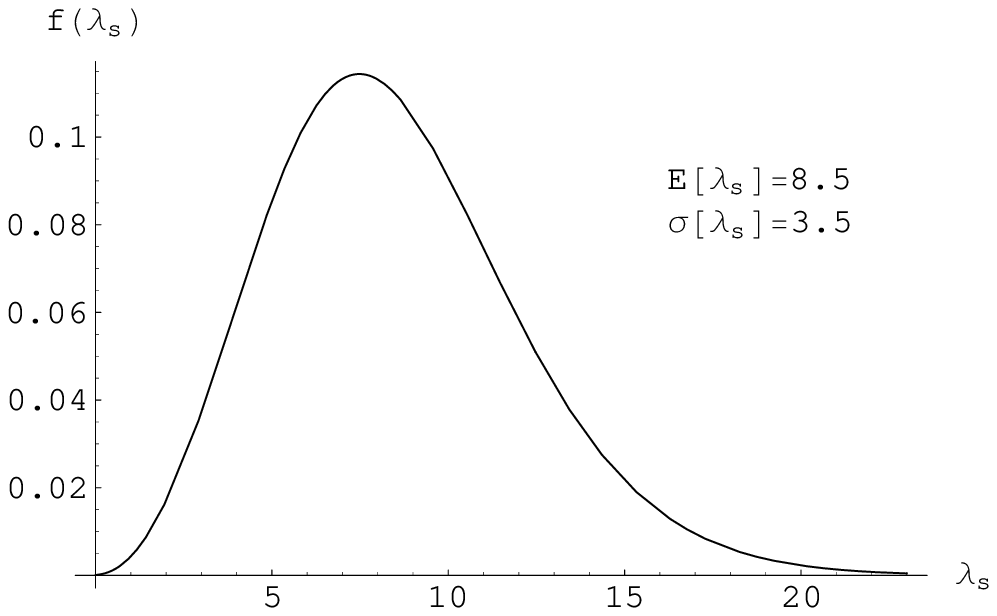,clip=,width=0.46\linewidth} \\
\hline
\end{tabular}
\end{center}
\caption{\small Inference about $n_s$ (histograms) 
and $p_s$ (continuous lines)
for $n=12$ and $x=9$,
assuming $\lambda_b=4$ and three values of
 $p_b$: 0.75, 0.25 and 0.95 (top down).}
\label{fig:bin_back_inf_ns}
\end{figure}

The histograms of 
Fig.~\ref{fig:bin_back_inf_ns} show examples of the probability
distributions of $n_s$ for $\lambda_b=4$ and three different
hypotheses for $p_b$. 
These distributions quantify how much we believe that $n_s$ 
out of the observed $n$ belong to the signal.
[By the way, the number $n_b$ of background objects 
 present in the data can be inferred as complement
to $n_s$, since the two numbers are linearly dependent. It follows
that $f(n_b\,|\,x,\,n,\,\lambda_b,\,p_b) = 
f(n-n_s\,|\,x,\,n,\,\lambda_b,\,p_b)$.]

\begin{figure}
\centering\epsfig{file=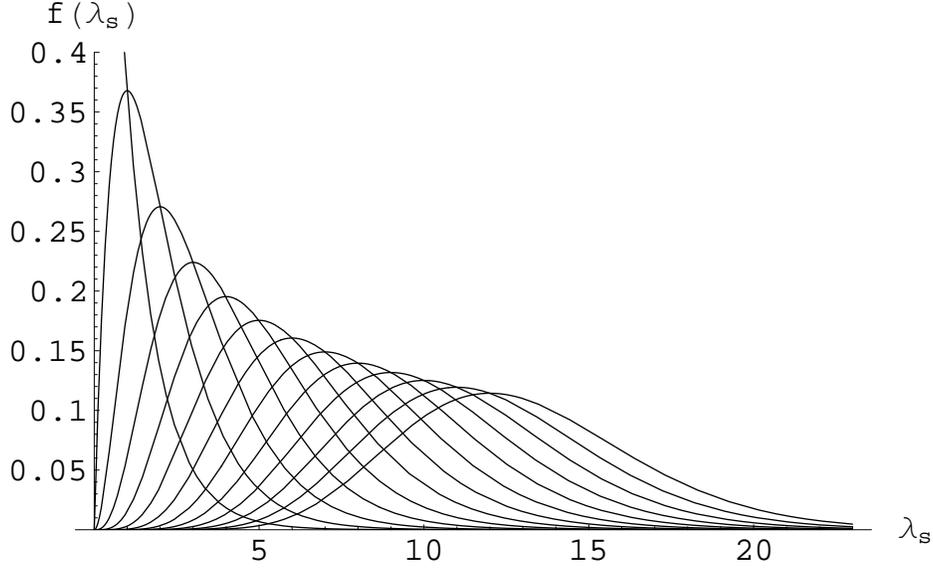,clip=,width=\linewidth}
\caption{\small Inference of $\lambda_s$ depending on the $n_s$,
ranging from 0 to 12 (left to right curves).}
\label{fig:inv_pois_0_12}
\end{figure}

A different question is to infer the  the Poisson $\lambda_s$
of the signal. Using once more Bayes theorem we get,
under the hypothesis of $n_s$ signal objects:
\begin{eqnarray}
f(\lambda_s\,|\,n_s) &\propto& f(n_s\,|\,{\cal P}_{\lambda_s}) \,.
f_0(\lambda_s)  
\end{eqnarray}
Assuming a uniform prior for $\lambda_s$ we get 
(see e.g. Ref.~\cite{BR}):
\begin{eqnarray}
f(\lambda_s\,|\,n_s) &=& \frac{e^{-\lambda_s}\,\lambda_s^{n_s}}{n_s!}\,, 
\end{eqnarray}
with expected value and variance both equal to 
$n_s+1$ and mode equal to $n_s$ (the expected value is shifted 
on the right side of the mode because the distribution is skewed 
to the right). 
Figure \ref{fig:inv_pois_0_12}
shows these pdf's, for $n_s$ ranging from 0 to 12 and
assuming a uniform prior for $\lambda_s$.

As far the pdf of $\lambda_s$ that depends on all possible 
values of $n_s$, each with is probability, is concerned, 
we get from  probability theory 
[and remembering that, indeed, 
$f(\lambda_s\,|\,n_s,\,x,\,n,\,\lambda_b,\,p_b)$ is equal to
 $f(\lambda_s\,|\,n_s)$, because $n_s$ depends only on 
$\lambda_s$, and then the other way around]:
\begin{eqnarray}
f(\lambda_s\,|\,x,\,n,\,\lambda_b,\,p_b) &\propto& 
\sum_{n_s} f(\lambda_s\,|\,n_s)\,f(n_s\,|\,x,\,n,\,\lambda_b,\,p_b)\,,
\end{eqnarray}
i.e. the pdf of $\lambda_s$ is the weighted 
average\footnote{It follows that all moments of the distribution
are weighted averages of the moments of the conditional distribution.
Then, expected value and variance of $\lambda_s$ can be 
easily obtained from the conditional expected values and variances:
\begin{eqnarray}
\mbox{E}(\lambda_s) &\propto& 
\sum_{n_s}\, \mbox{E}(\lambda_s\,|\,n_s)\,f(n_s)\, \nonumber \\
\mbox{Var}(\lambda_s) &\propto& 
\sum_{n_s}\, [\mbox{Var}(\lambda_s\,|\,n_s) +
\mbox{E}^2(\lambda_s\,|\,n_s)]\,f(n_s)\,. \nonumber
\end{eqnarray} 
} 
of the   several $n_s$ depending pdf's. 

The results for the example we are considering in this
section are given in the plots of Fig.~\ref{fig:bin_back_inf_ns}.

\section{Conclusions}
The classical inverse problem related to the binomial distribution
has been reviewed and extended to the presence of background
either only on the number of `successes', or on the trials 
themselves. The probabilistic approach followed here allows
to treat the problems only using probability rules.
The results are always in qualitative agreement with
intuition, are consistent with observations and prior knowledge
and, never lead to absurdities, like $p$ outside the range 
0 and 1. 

The role of the priors, that are crucial to allow the
probabilistic inversion and very useful to balance in the
proper way prior knowledge and evidence from new observations,
has been also emphasized, showing when they can be neglected
and when they are so critical that it is preferable not 
to provide probabilistic conclusions. 

\vspace{0.8cm}
It is a pleasure to thank Stefano Andreon for several
stimulating discussions on the subject.

\end{document}